\documentclass{ws-jbcb}
\usepackage{multirow}
\begin{document}

\markboth{Chignola, Del Fabbro, Farina, and Milotti}
{Computational challenges of tumor spheroid modeling}

%
\catchline{}{}{}{}{}
%

\title{COMPUTATIONAL CHALLENGES OF TUMOR SPHEROID MODELING}

\author{ROBERTO CHIGNOLA}

\address{Dipartimento di Biotecnologie, Universit\`a di Verona, and INFN -- Sezione di Trieste, \\ Strada le Grazie 15 - CV1, I-37134, Verona, Italia\\
roberto.chignola@univr.it}

\author{ALESSIO DEL FABBRO}

\address{Dipartimento di Fisica, Universit\`a di Trieste, and INFN -- Sezione di Trieste, \\
Via Valerio 2, I-34127, Trieste, Italia\\
delfabbro@ts.infn.it}

\author{MARCELLO FARINA}

\address{Dipartimento di Elettronica e Informazione, Politecnico di Milano, \\
Via Ponzio 34/5, I-20133, Milano, Italia \\
farina@elet.polimi.it}

\author{EDOARDO MILOTTI}

\address{Dipartimento di Fisica, Universit\`a di Trieste, and INFN -- Sezione di Trieste, \\
Via Valerio 2, I-34127, Trieste, Italia\\
milotti@ts.infn.it}

\maketitle

\begin{history}
\received{(Day Month Year)}
\revised{(Day Month Year)}
\accepted{(Day Month Year)}
\end{history}

\begin{abstract}
The speed and the versatility of today's computers open up new opportunities to simulate complex biological systems. Here we review a computational approach recently proposed by us to model large tumor cell populations and spheroids, and we put forward general considerations that apply to any fine-grained numerical model of tumors. We discuss ways to bypass computational limitations and discuss our incremental approach, where each step is validated by experimental observations on a quantitative basis. We present a few results on the growth of tumor cells in closed and open environments and of tumor spheroids. This study suggests new ways to explore the initial growth phase of solid tumors and to optimize anti-tumor treatments.
\end{abstract}


\section{Introduction}

\subsection{Tumor growth as a multi-scale biological process}
\noindent
Biological systems span vast spatiotemporal scales, from the microscopic dynamics of atoms to the macroscopic dynamics of cell clusters. Information flows in both directions and determines the behavior of living matter and ultimately the normal physiology of organisms and the onset of pathologies such as tumors. Individual tumors are complex biological systems and, in spite of great therapeutic advances, many tumors still escape treatment and lead to death. Indeed, the malignancy and the response of tumors to therapy depends on their growth potential which in turn is determined by the ability of tumor cells to adapt to different environments, to compete with normal cells for both space and nutrients and to ignore molecular signals attempting to block cell cycling or to promote cell death.\cite{r1} Major efforts have been made by experimental scientists to highlight the molecular circuits, 
that are often altered as the consequence of genetic alterations, underlying tumor cell biology and, on this basis, to develop novel therapeutic strategies. This has produced a huge body of knowledge which has deepened our understanding of the molecular details of tumor cell biology but often with little or no consequence for the clinical management of tumors. Tumor resistance to radiotherapy and chemotherapy is in fact the main reason for ineffective therapy, and this lack of responsiveness is only partly 
caused by genetic alterations of tumor cells. Non genetic factors, such as pH, ions other than H$^{+}$ and oxygen distributions, nutrient supply, proteins of the extracellular matrix, that constitute the tumor microenvironment, have also been found important. Indeed, microenvironmental changes can protect tumor cells from apoptosis, promote their survival and the progression of both epithelial tumors and non-solid hematologic malignancies. \cite{r1A,r1B}\\
Unfortunately, it is very difficult to investigate the dynamics of  microenvironmental changes in real tumors, as this would require non-invasive time-resolved technologies with appropriate spatial and temporal resolutions, and even if these technologies were available it would be extremely complicated to explore competing hypotheses of
tumor microevolution, to detail and control the underlying mechanisms and to predict the impact of this knowledge on the clinical management of tumors.\\ 
At least part of the complexity of the problem is a sheer consequence of tumor size: clinicians deal with the macroscopic properties of tumors, i.e. masses that may eventually weigh a few kilograms, and thus with a number of cells that ranges between $10^6$ and $10^{13}$, and that may grow for months or years, with a corresponding number of cell cycles somewhere in the range between 100 and 10000. Moreover, at the microscopic level the malignant transformation of single cells is a multistep process that involves the modification of several molecular circuits which, in turn, modify the cell's behavior and the relationships between cells and the environment.\cite{r1} In addition, the epigenetic and environmental factors, which include cell-cell interactions, also conspire with genetic information to make tumor growth a highly variable process with very strong feedbacks.\cite{r1,r1A} The highly nonlinear character of the cells' internal molecular machinery, combined with the cell-cell and environmental interactions, with the large number of cells in a tumor, and with the extended tumor lifespan, make predictions based on the behavior of individual molecular circuits utterly haphazard.\\ 
The availability of powerful computers has already helped bridge the gap between observations and predictions in many complex problems, and this suggests that in the future we shall be able to simulate the behavior of large cell populations {\it ab initio}, starting from individual molecular reactions in single cells and climbing the ladder of complexity up to the behavior of whole multicellular organisms. This requires an incremental approach, a ladder of increasingly complex models of cells, bearing in mind that each step must be validated by experimental observations. 

\subsection{Multicell tumor spheroids: an in vitro cell model with intermediate complexity between real tumors and conventional tumor cell cultures}
\noindent
Small tumor cell aggregates (volume $\le 1 \mathrm{mm}^3$) may escape conventional treatment of solid tumors and, in time, may lead to recurrence of the primary pathology, often with a different phenotype (e.g., acquired resistance to chemotherapy, acquired ability to metastatize, etc.);\cite{r2,r3} cells can also  grow up to masses of this size without the support of the vasculature, although recent work is challenging this traditional view.\cite{r4}\\
Unfortunately it is very difficult to study these micromasses both in humans and in animal models, because their size is below the imaging limit of current technologies and it is not possible to measure their biological parameters and obtain information to validate the results of numerical simulations and draw conclusions on their biological and clinical properties.  Multicellular tumor spheroids represent a valid and effective experimental cell culture technique capable of preserving the three-dimensional topology of actual tumor cell clusters.\cite{r5,r6} Indeed, it is the three-dimensional topology that determines many important biological features, like the expression of specific genes, a slowed-down diffusion of nutrients and waste, and also the expression of new phenotypes like the resistance to radiotherapy, and in fact multicell tumor spheroids display many interesting biological properties that cannot be observed in monolayer cultures such as:\cite{r5,r6}
\begin{itemlist}
 \item heterogeneous expression of membrane receptors (that regulate cell adhesion and metabolism and also may act as target for specific anti-tumor drugs);
 \item production of an intercellular matrix (important for cell aggregation and for penetration of cells of the immune system);
  \item heterogeneous distribution of nutrients and oxygen that lead to the formation of a necrotic core and to a gradient of cell proliferation;
  \item appearence of resistance phenomena and/or heterogeneous response to antitumor therapies;
  \item growth kinetics very similar to those observed {\it in vivo}.
\end{itemlist}
Multicell tumor spheroids are thus intermediate between traditional cell cultures and tumors {\it in vivo}, and at the same time they are accessible to experimental measurements: they provide many data that can be used to test and validate multi-scale models of solid tumor growth in the prevascular phase. They are morphologically similar to small tumors below the detection threshold, and they share with them the lack of vascularization. For these reasons tumor spheroids are the perfect targets for mathematical and computational models of tumor growth.

\subsection{Tumor spheroids as targets for mathematical and computational models}
\noindent
Several mathematical and computational models of solid tumors and of spheroids have been developed since the pioneering work of Burton,\cite{r6A} and most of them have singled out some important feature of tumors,  such as the diffusion gradients of pH, oxygen and nutrients, or more recently, the formation of new blood vessels and/or the migration of tumor cells out of the primary lesion. It is not the aim of this work to provide a comprehensive review of the many models developed so far, since this has been the subject of recent survey papers.\cite{r6B,r6C} We note, however, that:
\begin{itemlist}
\item most models (with few exceptions like, e.g., refs.\cite{r6D,r6E,r6E1}), assume arbitrary units and outputs cannot be compared to actual experimental data: they describe only qualitative behaviors that somehow resemble those displayed by real avascular tumors. Due to the nonlinear character of the underlying biological processes, and hence of model equations, many different models can in principle produce outputs describing the same qualitative behaviors, and 
it is not possible to validate/falsify such models;
\item the appearance of widely different resistance phenomena to antitumor therapies in similarly grown, isolated, tumor spheroids of the same cell type\cite{r6F} indicates that random fluctuation phenomena play an all-important role in the growth kinetics of spheroids. It is well-known that the discrete events at the single-cell level do display some randomness (e.g., mitosis, cell death, partitioning of sub-cellular organelles and molecules between daughter cells at division, etc.) and one can pinpoint the source of large-scale variability on these fluctuations, as they are amplified and propagated by cell-cell and cell-environment interactions;
\item to the best of our knowledge, experimental evidence on the migration of labelled cells and microspheres into tumor spheroids\cite{r6G,r6G1} -- that reveal the existence of convective cell motions in the spheroid structure -- have not yet been fully explained by mathematical models in spite of various attempts;\cite{r6B}
\item in real tumor spheroids, cells communicate with other cells and the environment. In particular, concentration gradients of several molecular species, facilitated transport processes into and out of individual cells, and the mechanical forces that push and pull cells as they proliferate with repeated mitoses and then shrink after death, play an extremely important role in influencing the biochemical/biomechanical aspects of spheroid growth, and they depend in turn on the structure of the extracellular space. These processes mix with complex nonlinear interactions between the biochemical and the biomechanical part;  
\end{itemlist}

The absence of vascularization in multicell tumor spheroids and their near sphericity hide an internal complexity which is not easy to tame either with analytic models, or with numerical models based on rough simplifications of the biological settings such as cellular automata or lattice-based models. New modeling approaches {\it should} 
therefore be developed, and these should attempt to reproduce the biological complexity shown by real spheroids such as the interaction of cells with their neighbors and with the environment. Nonlinearity and discreteness of processes at the single-cell level suggest that the search for 
``logical rules'' {\it with simple grid-based models will not be sufficient, and that models that are capable of an equivalent level of complexity may be required.} 
Indeed, our final goal is the simulation of the microscopic interactions among cells and their environment to understand how these could affect spheroid growth and response to treatment.

\subsection{Complexity issues in real-life modeling approaches of cell populations}
\noindent
In principle, an all encompassing simulation could start with the atoms in the cell: using the methods of molecular dynamics it would then be possible to simulate a living being starting from atoms, molecules, and a description of the forces that bind them.\cite{r7} Unfortunately, at present this is impossible.\\
We can clarify the complexity issue by considering the problem of exploding memory size. If we take a cell radius $\sim 5\; \mu \mathrm{m}$, then the cell volume is approximately $5 \times 10^{-16} \;\mathrm{m}^3$, and the corresponding cell mass is about $5 \times 10^{-10} \mathrm{g}$, and this means that a single cell corresponds to about $10^{13}$-$10^{14}$ atoms. On the other hand, if we take a spatial resolution 0.01 nm (approximately one tenth of the diameter of a hydrogen atom), and aim to simulate a system size of 1 mm, then for each coordinate we need at least  a 24-bit dynamic range (3 bytes per coordinate), and thus at least 19 bytes/atom (3 coordinates plus 3 velocities plus one atom label), and about $10^{14}$-$10^{15}$ bytes/cell. Finally, the full simulation of a $10^6$ cell spheroid would require at least $10^{20}$-$10^{21}$ bytes/spheroid, and we see that present-day computers are pitifully inadequate for such a brute force approach.\\
Likewise, the time complexity of simulation algorithms would also be unmanageable: here we must assume that we are somehow able to tame the $O(N^2)$ complexity of binary interactions between the simulation elements and also the complexities of several subalgorithms like matrix inversion and the like, and that the overall algorithmic time-complexity of the simulation program adds up to a mere $O(N)$. Since the fastest dynamics in acqueous solutions is determined by the motion of protons in the hydrogen bonds in water,\cite{r8} and is of the order of 1 ps, and since there are approximately $10^{13}$ hydrogen bonds in each cell, then one must take at least $10^{25}$ time steps just to simulate 1 s of a single
 proton motion in a cell (and with a rather poor time resolution).\\
These approximate calculations amount to an operational definition of biological complexity, and they show that at the moment we cannot even dream of carrying out a true {\it ab initio} simulation of tumor spheroids. Therefore -- in our approach -- we have chosen the highest level scale at which the complexity of individual molecular processes is still somehow evident, the cell scale, although the question remains whether it is possible to stay at this level and still 
provide a comprehensive and realistic description of cell behaviour. 

\section{Models}

\subsection{A minimal model of the tumor cell}
\noindent
When viewed at the mesoscopic scale the intracellular molecular machinery displays an astonishing complexity, with a huge number of intertwined chemical reactions that mark the different phases of the cell's life. On the other hand, a biophysical simulator of tumor spheroids must start from a realistic description of the tumor cell, and this ultimately means that at least cell metabolism and its interconnection with the cell cycle must be modeled at a sufficient level of detail in order to describe how the behavior of a single cell is affected by the other cells in the cluster, by the chemical composition of the environment and by physical parameters such as temperature, density, radiation, etc.\\
Our approach is based on the fact that biochemical networks in the cell possess a hierarchical structure.\cite{r9} It is known that, if a network has such topology, then the system dynamics are dominated by the network's hubs.\cite{r9} Thus, by modeling the hubs of the cell's biochemical network one should, at least in principle, be able to capture most of the information of the cell's biochemical dynamics.\cite{r10,r11} This also means that we must necessarily parameterize and average many details of cell metabolism. In this way, we achieve a significant reduction in computational complexity and a considerable reduction of the space-time scale problems that affect simulations aimed at calculating the properties of macroscopic objects starting from microscopic models.

\begin{figure}[th]
\centerline{\psfig{file=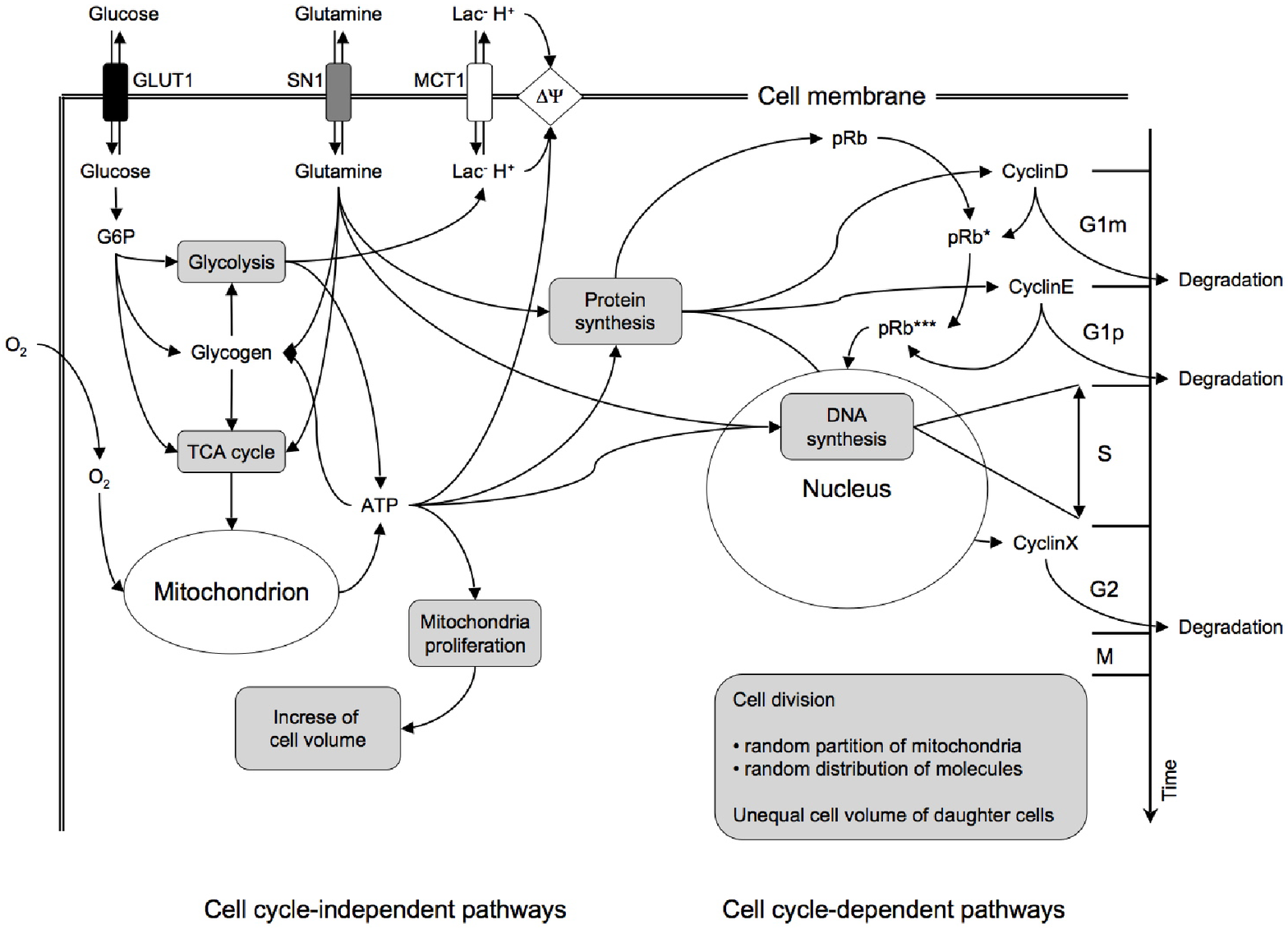,width=13cm}}
\vspace*{8pt}
\caption{Metabolic network implemented in the simulation program. See text for details. A complete description of this network can be found in reference 22}
\end{figure}

Figure 1 shows a sketch of the metabolic pathways that have been modeled so far.\cite{r10,r11}
We have taken into consideration both cell cycle-independent and cell cycle-dependent pathways. Among the former are the uptake and utilization of nutrients such as glucose and glutamine and the elimination of waste molecules such as lactate. This part describes quantitatively the production of ATP by both oxydative phosphorylation and glycolysis and it bridges the two paths: in fact, ATP and glutamine constitute the building blocks of proteins and DNA\cite{r13,r14,r15} and their availability drives the kinetics of protein and DNA synthesis. These in turn determine the expression of proteins that regulate the cell cycle such as pRb and cyclins and the duration of specific phases of the cell cycle such as the S phase.\cite{r16} Proteins that regulate the cell cycle also determine the duration of the other phases, and we have implemented a thresholding mechanism based on multisite phosphorylation to determine when a cell steps beyond cell cycle checkpoints.\cite{r17,r18}\\
ATP availability also determines the proliferation of mitochondria, a biological process that has been coupled phenomenologically to the increase of the cell volume, while ATP deprivation leads to cell death -- although the model also takes into account several other mechanisms of cell death.\cite{r11} Finally, we take into consideration stochastic aspects such as the random partitioning of mitochondria and the random distribution of molecules at cell division, that contribute to desynchronize the duration of the cell cycle in daughter cells.\cite{r10,r11}\\
The rationale for sorting the above processes out of the complex metabolic machinery of the cell has been discussed in detail in our previous papers.\cite{r10,r11} Because of the interconnections among the pathways, we believe that this model cannot be further reduced, above all if we are to produce a realistic simulation of how cells exchange biochemical information with the environment and the other cells, and regulate their life cycle.

\subsection{Modeling tumor spheroids}
\noindent
The metabolic model of tumor cells shown in figure 1 is the basis on which we build the simulation of tumor spheroids and, for validation purposes with experimental data, it has been used to simulate the growth of large cell populations in both closed and open environments.\cite{r10,r11} The simulations of dispersed cells have been successful,\cite{r10,r11} however the modeling of tumor spheroids requires several other biological, chemical and physical aspects:\cite{r19,r20}
\begin{romanlist}[(ii)]
 \item adequate numerical modeling of diffusion processes. The diffusion of chemicals in a cell cluster proceeds either by normal diffusion or by facilitated diffusion across cell membranes. Facilitated diffusion is mostly a biochemical process that has a weaker dependence on concentration gradients and is brought about essentially by transporters expressed at the cell surface. For facilitated diffusion to proceed in cell clusters, it is necessary to include the extracellular spaces as basic elements.  Neighboring cells modify the surrounding environment and compete for the same resources, and thus it is important to define the proximity relationships among cells in the cluster. This is accomplished by a specialized part of the program (see below);
 \item  environment. The environment is also taken into account in the simulation: it is defined as the external volume of nutrient fluid that communicates by diffusion with the extracellular spaces that surround cells, and it is modeled as a single compartment. The environment is modified both by the fast diffusion processes that transport nutrients and metabolites into and out of the cell cluster, and, in case, by the slow flushing of nutrient fluid in a bioreactor enclosure;
 \item biomechanical interaction. In a cluster, cells interact mechanically as well as biochemically: this part of the simulation program is essentially a simple integrator like those found in dissipative dynamics simulations. Cells are approximated by soft spheres that move in a highly viscous environment;\cite{r19}
 \item geometry. Both the mechanical and the diffusion part of the program require a knowledge of the proximity relationships among neighboring cells. We wish to point out that, unlike other models, in our simulation program cells do not grow in an environment defined by a fixed grid, where it is difficult to model appropriately processes such as cell division, and that cells are free to move in the three-dimensional space as as they are pushed and pulled by forces resulting from biomechanical interactions between cells. The nearest neighbors are defined by the links in a Delaunay triangulation\cite{r21} and they are computed by the triangulation methods in the computational geometry package CGAL (see http://www.cgal.org). In this way all the computational complexity of binary interactions is reduced from a potential $O(N^2)$ to a much more manageable $O(N)$.
\end{romanlist}

A full description of our model that includes all technical details can be found in reference 32.

\subsection{Stability issues in the numerical integration of model differential equations}
\noindent
In our model each cell is described by a metabolic network and by other mechanisms that include both discrete deterministic and stochastic events. The description is thus mixed, with smooth evolutions interspersed with discrete steps. The exchange of molecules with the surrounding environment means that transport into and out of cells is closely linked with diffusion processes that involve the whole cluster of cells, and finally lead to a very large set of differential equations. Since our goal is to simulate tumor spheroids up to a diameter of 1 mm, which corresponds to about 1 million cells, the software must eventually solve a very large number of coupled nonlinear equations, as many as $10^7$-$10^8$ equations (because there are -- at present -- 
25 variables per cell). These equations are similar to other equations encountered in systems biology, but the number is uncommonly large. In addition, the equations are quite stiff, since they describe processes that range from fast diffusion in small extracellular spaces (a few tens of $\mu$s) up to the slow development of the spheroid as a whole ($\sim 10^{7}\; \mathrm{s}$) and thus the characteristic times span about 12 orders of magnitude. We have solved the complex stability problems that arise in such a situation\cite{r22} and thus this model now stands up as a true multiscale model, both in space and, even more so, in time. We argue here that a modeling such as that described in reference 33  is mandatory in any such simulation of cell clusters.\\
We also remark that some of the model parameters slowly change as cells grow and this is once again at variance with most differential systems used in systems biology where parameters are fixed. Finally, the continuous system evolution described by the differential system is interrupted at random times by discrete events; these events may be internal transition in individual cells (in this case the system parameters change abruptly from one integration step to the next) or cellular division events (in this case the number of equations changes). For all these reasons, we believe that our simulation program is unique in the sense that it tackles simultaneously and  for the first time a vast array of technical issues that are not addressed by any other model. In particular, the same implementation scheme of model equations can be used to simulate both fast and slow processes simply by tuning the integration time step, and this is an added value to the simulation program that might be exploited to investigate at will biological events with different characteristic times.   

\section{Results}

\subsection{Simulations of the growth of dispersed cells in a closed environment}
\noindent
One important aspect of the approach described in the previous section is that it is fully quantitative and its outputs can be directly compared with available experimental data.
The model of cell metabolism, growth and proliferation described above can be used to simulate a population of dispersed cells growing in a closed environment. This is equivalent to considering cultures of blood cells in vitro, such as leukemia cells. Cells growing in a closed environment establish a sort of negative feedback with the environment itself. While cells grow, they consume nutrients and release waste molecules that acidify the medium. As the environment gradually becomes more and more acidic, the uptake of nutrients is also reduced and can eventually switch off completely, thereby leading to a depletion of the energy reserves and ultimately to cell death. This mechanism involves the whole model of cell metabolism and control of the cell cycle and can be tested experimentally because it defines the carrying capacity of the environment where cells are grown.\\
We have already demonstrated that our model nicely reproduces common growth curves observed {\it in vitro}\cite{r11} and shows predictive capabilities, in that it can also describe the growth of leukemia cells under non conventional conditions, where, e.g., the biochemical composition of culture media is periodically refreshed. The model predictivity in nonconventional conditions is obviously an important test, because it demonstrates that the model is not just a qualitative description, but can be used to set up a true {\it in silico} laboratory.
Indeed, the model outputs are in good quantitative agreement with actual data on metabolic and cellular parameters (Table 1)\cite{r10,r11}.

\begin{table}[th]
\tbl{Estimated morphologic, kinetic and metabolic parameters for a population of dispersed tumor cells and comparison with actual experimental data.\label{t1}}
{\begin{tabular}{@{}lccc@{}} \toprule
Parameter & Simulated & Experimental & Reference \\
  \colrule
\it{Morphologic} & \multicolumn{1}{c}{\it{Average}\hspace{3pt} \it{Min}\hspace{3pt} \it{Max}}\\

Radius ($\mu$m) & \multicolumn{1}{c}{\hphantom{00}5.0\hspace{16pt}4.8\hspace{12pt}5.3} & 5.5 -- 7.1 & \cite{r26}\\
Volume($\mu$m$^{3}$) & \multicolumn{1}{c}{\phantom{00}530\hspace{14pt}471\hspace{8pt} 623} & 700 -- 1500 & \cite{r26,r27}\\
Mitochondria/cell & \multicolumn{1}{c}{\phantom{00}220.4\hspace{8pt}190.6\hspace{2pt} 266.9} & \hphantom{0}83 -- 677\tabmark{a} & \cite{r28}\\
\\
\it{Kinetic}\\
Growth rate\tabmark{b} (h$^{-1}$) & \multicolumn{1}{c}{\hphantom{00}0.035\hphantom{0000000000000}} & \hphantom{0}0.03 -- 0.035\tabmark{c} & \cite{r11}\\
Doubling time\tabmark{b} (h) & \multicolumn{1}{c}{\hphantom{00}19.8\hphantom{0000000000000}} & 19.7 -- 22.8\tabmark{c} & \cite{r11}\\
G1 ($\%$) &  \multicolumn{1}{c}{\hphantom{00}52.5\hspace{12pt}48.4\hspace{8pt} 59.3} & 54.4 $\pm$ 2.2\tabmark{c} & \cite{r10}\\
S ($\%$) &  \multicolumn{1}{c}{\hphantom{00}34.5\hspace{12pt}30.5\hspace{8pt} 40.5} & 27.5 $\pm$ 5.8\tabmark{c} & \cite{r10}\\
G2/M ($\%$) &  \multicolumn{1}{c}{\hphantom{00}12.9\hspace{14pt}7.3\hspace{10pt} 17.7} & 16.4 $\pm$ 1.7\tabmark{c} & \cite{r10}\\
\\
\it{Metabolic}\\
ATP/cell\tabmark{d} & \multicolumn{1}{c}{\hphantom{00}5.5\hspace{16pt}5.4\hspace{12pt} 5.6} & 4.3 -- 5.8 & \cite{r10}\\
Glucose uptake\tabmark{e} & \multicolumn{1}{c}{\hphantom{000}1.9 $\pm$ 0.3\hphantom{0000000000000}} & 2.5 $\pm$ 0.2 & \cite{r29}\\
Lactate production\tabmark{e} & \multicolumn{1}{c}{\hphantom{000}3.8 $\pm$ 0.3\hphantom{0000000000000}} & 3.9 $\pm$ 0.8 & \cite{r29}\\
ATP production\tabmark{e,f} & \multicolumn{1}{c}{\hphantom{000}19.8 $\pm$ 8.3\hphantom{0000000000000}} & 37.8 & \cite{r29}\\
ATP production\tabmark{e,g} & \multicolumn{1}{c}{\hphantom{000}10.6 $\pm$ 1.3\hphantom{0000000000000}} & 11.4 $\pm$ 2.3 & \cite{r29}\\
Oxygen consumption\tabmark{e} & \multicolumn{1}{c}{\hphantom{000}0.25 $\pm$ 0.1\hphantom{0000000000000}} & 0.48 $\pm$ 0.1 & \cite{r29}

\\ \botrule
\end{tabular}}
\begin{tabfootnote}
\tabmark{a} Range of the number of mitochondria observed in different cell types\\
\tabmark{b} The growth rate for both simulated and experimental cell populations was calculated by exponential fitting of growth curves. The doubling time was then calculated as $\log 2/(\mathrm{growth\; rate})$\\
\tabmark{c} Data measured for MOLT3 (human T lymphoblastoid cell line) and Raji (human B lymphoblastoid cell line) cells in our own experiments\\
\tabmark{d} Values are expressed as 10$^{-18}$ kg\\
\tabmark{e} Values are expressed as 10$^{-19}$ kg s$^{-1}$\\
\tabmark{f} ATP production through oxidative phosphorylation\\
\tabmark{g} ATP production through glycolysis
\end{tabfootnote}
\end{table}

\subsection{Simulations of the growth of dispersed cells in an open environment}
\noindent
The negative feedback between cells and their environment discussed above can be partially removed by opening up the environment, setting up a flow that removes consumed culture medium and replaces it with fresh medium. Experimentally, this condition is realized in bioreactors such as those used to culture cells at high density for biotechnology purposes, e.g., antibody production.\cite{r23} Under these conditions, viable cells are expected to reach a steady state given by the dynamic equilibrium between proliferation and death.\\ 
Figure 2 shows the result of a simulation campaign carried out with the numerical model in reference 22.  Parameter values were left unchanged with respect to previous simulations, and the only difference is that here a continuous flux of fresh medium slowly replaces the culture medium with rates comparable to those in actual experiments with bioreactors. As we see in figure 2, there is a good correspondence between simulations and actual data for different flow rates.\\
This kind of virtual experiment was not planned during model development and model parameters were not tuned to take into account the growth of simulated cells in an open environment. This further demonstrates the predictive power of the model and its potential use as an {\it in silico} cell biology laboratory. 

\begin{figure}[th]
\centerline{\psfig{file=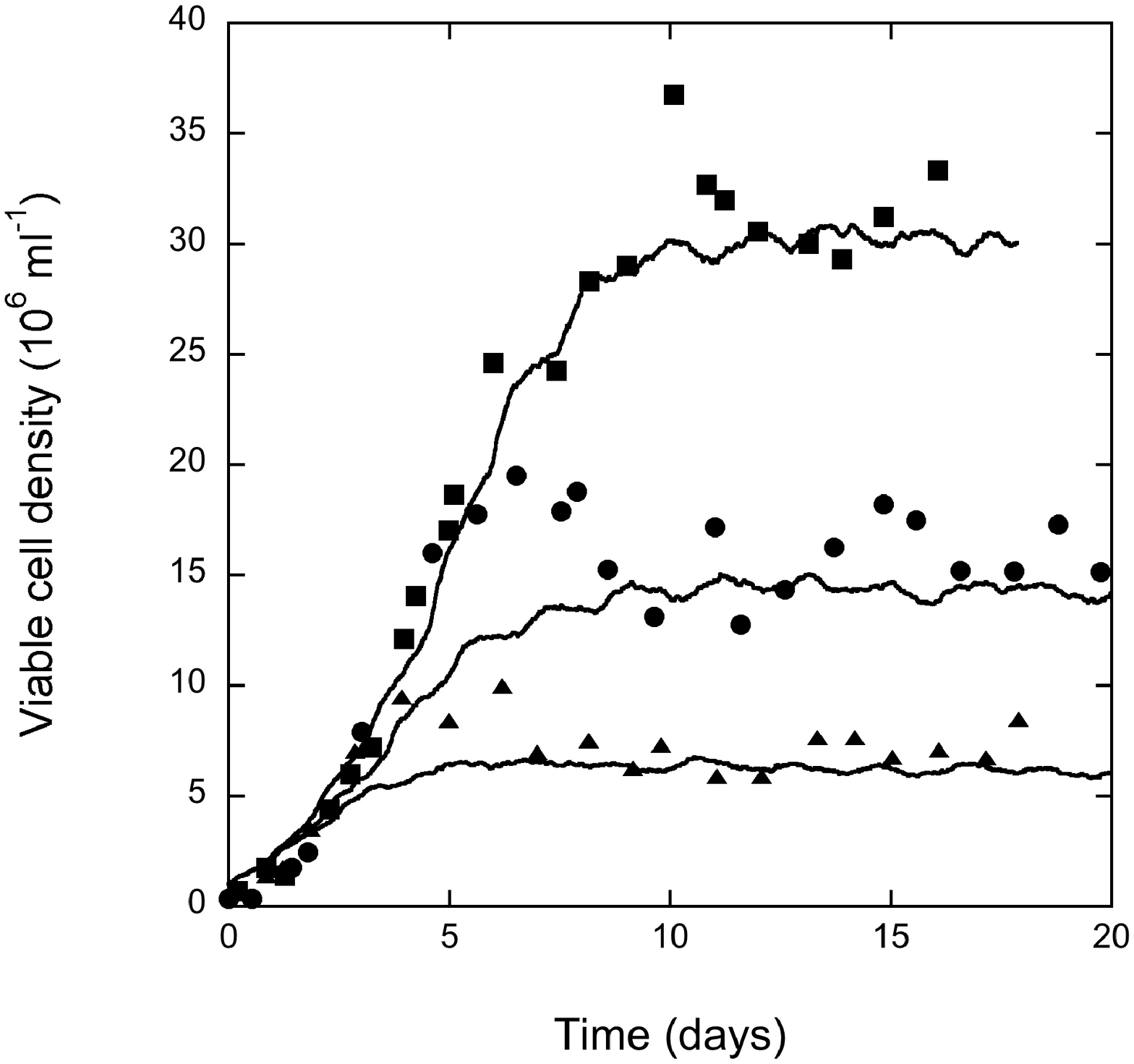,width=10cm}}
\vspace*{8pt}
\caption{Growth kinetics of dispersed cells in a bioreactor. Points represent experimental data taken at different flux rates of fresh medium and are: squares, 2.33 volumes of fresh medium perfused / effective suspension volume / day (vvd); circles, 1 vvd; triangles, 0.48 vvd. Lines represent simulation outputs obtained with our program for the same fluxes. See also the text for further details. Experimental data are taken from reference 34.}
\end{figure}

\subsection{Climbing the third dimension: a simulation test run of the growth of tumor spheroids}
\noindent
Our model can be used to simulate the behavior of large tumor cell populations growing both in closed or in open environments. The very same model can also be used to simulate the growth of avascularized tumor spheroids, where we have to take into account additional biological, chemical, physical and mathematical aspects. 
Cells in a spheroid adhere to one another. Thus, one must include and model biomechanical forces that act upon cells. Cells have been modeled as soft spheres interacting through visco-elastic forces.\cite{r19,r21A} These forces allow the whole cell aggregate to preserve a three-dimensional structure in spite of the push that cells exert at mitosis on neighboring cells 
as they compete for space. This results in a dynamic balance between repulsive and adhesion forces that shapes the spheroid structure in the course of its development.\\ 
Each cell is surrounded by a small free volume that corresponds to the extracellular space. This is fundamental to allow nutrients and waste molecules to diffuse freely through the cell cluster. The inclusion of diffusion, however, introduces processes that occur with very short characteristic times, in the order of a few tens of $\mu\mathrm{s}$, and this increases the stiffness of the underlying system of differential equations. 

\begin{figure}[th]
\centerline{\psfig{file=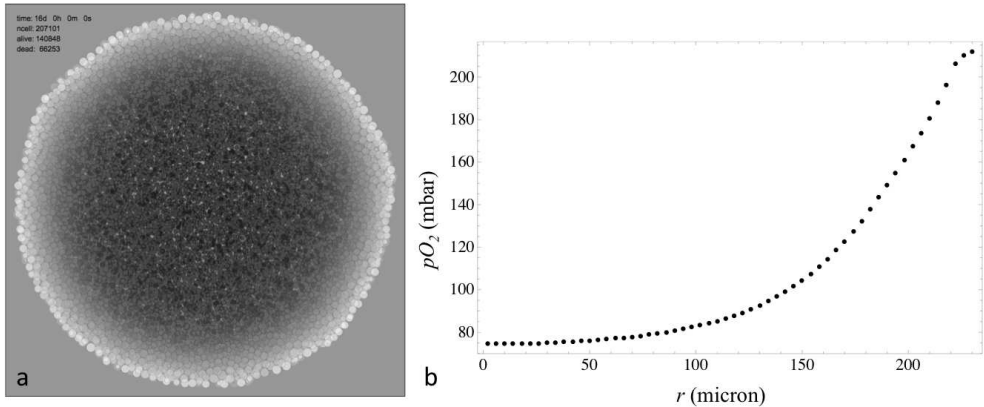,width=13cm}}
\vspace*{8pt}
\caption{Oxygen concentration profile in a simulated spheroid after 16 days of simulated time. Panel a: slice cut through the center of the spheroid showing the oxygen concentration. The gray levels map the concentration value: white, higher concentration; black, lower concentration (the background is light gray to provide suitable contrast). Panel b: oxygen concentration given as atmospheric partial pressure vs. the distance $r$ from the centroid of the simulated tumor spheroid (the rightmost value corresponds to the standard atmospheric partial pressure).}
\end{figure}

Figure 3 shows a result obtained in a run with a simulated spheroid of more than 200000 cells and 16 days old (simulated time). The oxygen concentration decreases towards the center of the spheroid and one can easily detect a spatial gradient that is very similar to that observed in real experiments. As the spheroid grows the internal environment become hypoxic, and this contributes to cell death and to the formation of a necrotic core.\\ 
Once again, we stress that our approach is quantitative and this allows us to test model outputs with actual experimental data. The profiles of the main metabolic actors have been measured in spheroids and data on at least some cell and morphologic parameters are also available for comparison. In Table 2 we show that our model of multicell tumor spheroids provides outputs that compare favorably with available microscopic data, and in Figure 4 we show that the growth kinetics of spheroids can also be properly simulated.

\begin{table}[th]
\tbl{Estimated metabolic, histologic and kinetic parameters for a virtual multicell tumor spheroid and comparison with actual experimental data.\label{t2}}
{\begin{tabular}{@{}lccc@{}} \toprule
Parameter & Simulation & Experiments & References \\
  \colrule
\it{Metabolic}\\
Glucose uptake\tabmark{a} (kg s$^{-1}$ m$^{-3}$) & 1.44 $\cdot$ 10$^{-3}$ & 5.4 -- 12.6 $\cdot$ 10$^{-3}$ & \cite{r31}\\
Lactate release\tabmark{a} (kg s$^{-1}$ m$^{-3}$) & 1.35 $\cdot$ 10$^{-3}$ & 5.4 -- 9 $\cdot$ 10$^{-3}$ & \cite{r31}\\
pO$_2$\tabmark{b} (mmHg) & 7 & 0 -- 20 & \cite{r31}\\
pH\tabmark{c} & 6.7 & 6.6 -- 6.99 & \cite{r32,r33}\\
$\Delta$pH\tabmark{d} & 0.77 & 0.49 $\pm$ 0.08 & \cite{r33}\\
\\
\it{Histologic} \\
Viable cell rim thickness\tabmark{e} ($\mu$m) & 155 & 142 -- 310 & \cite{r33,r34,r35}\\
Hypoxic rim thickness\tabmark{f} ($\mu$m) & 98 & 44 $\pm$ 52 & \cite{r35}\\
\\
\it{Kinetic}\\
Cell cycle distribution\\
G1 ($\%$) & 57.3 & 58 $\pm$ 4 & \cite{r36}\\
S ($\%$) & 21.6 & 19 $\pm$ 1 & \cite{r36}\\
G2/M ($\%$) & 21.1 & 23 $\pm$ 1 &\cite{r36}
\\ \botrule
\end{tabular}}
\begin{tabnote}
Metabolic and histologic parameters in spheroids of approximately 500 $\mu$m diameter.
\end{tabnote}
\begin{tabfootnote}
\tabmark{a} Rate of glucose uptake or lactate release per viable spheroid volume\\
\tabmark{b} Central pO$_2$ tension (experiments) or estimated in the centroid (simulations)\\
\tabmark{c} pH has been determined in the central region of the spheroids. This corresponds to a sphere radius $\approx$ 100 $\mu$m about the centroid of the spheroid \\
\tabmark{d} Difference between environmental pH and pH 200 $\mu$m below the spheroid surface\\
\tabmark{e} In our simulations the viable cell rim thickness corresponds to the distance between the spheroid surface and the inner shell where only 5 $\%$ of the cells are still alive\\
\tabmark{f} These values corresponds to the radius of the necrotic core\\
\end{tabfootnote}
\end{table}

Thus, our model connects the mesoscopic and the macroscopic scales. It is also important to note that the data in Figure 3, Figure 4 and Table 1 do not constitute a final validation of our model and more virtual experiments must be carried out, possibly under conditions that were not considered during model development. These include the growth of spheroids in media with a different concentration of glucose and oxygen. Experimental data are available\cite{r23A}, and virtual experiments can be carried out because of the versatile model of cell metabolism that we have implemented in our model of multicell tumor spheroids.\\ 
Since concentration profiles are computed during the time evolution of the whole cluster, the dynamic variation of the distributions of molecular species can also be appreciated and studied for the first time. This computational approach could eventually help to improve our knowledge on the initial growth phase of avascular solid tumors, as it discloses details that are otherwise very difficult to detect and measure.\cite{r21A}

\begin{figure}[th]
\centerline{\psfig{file=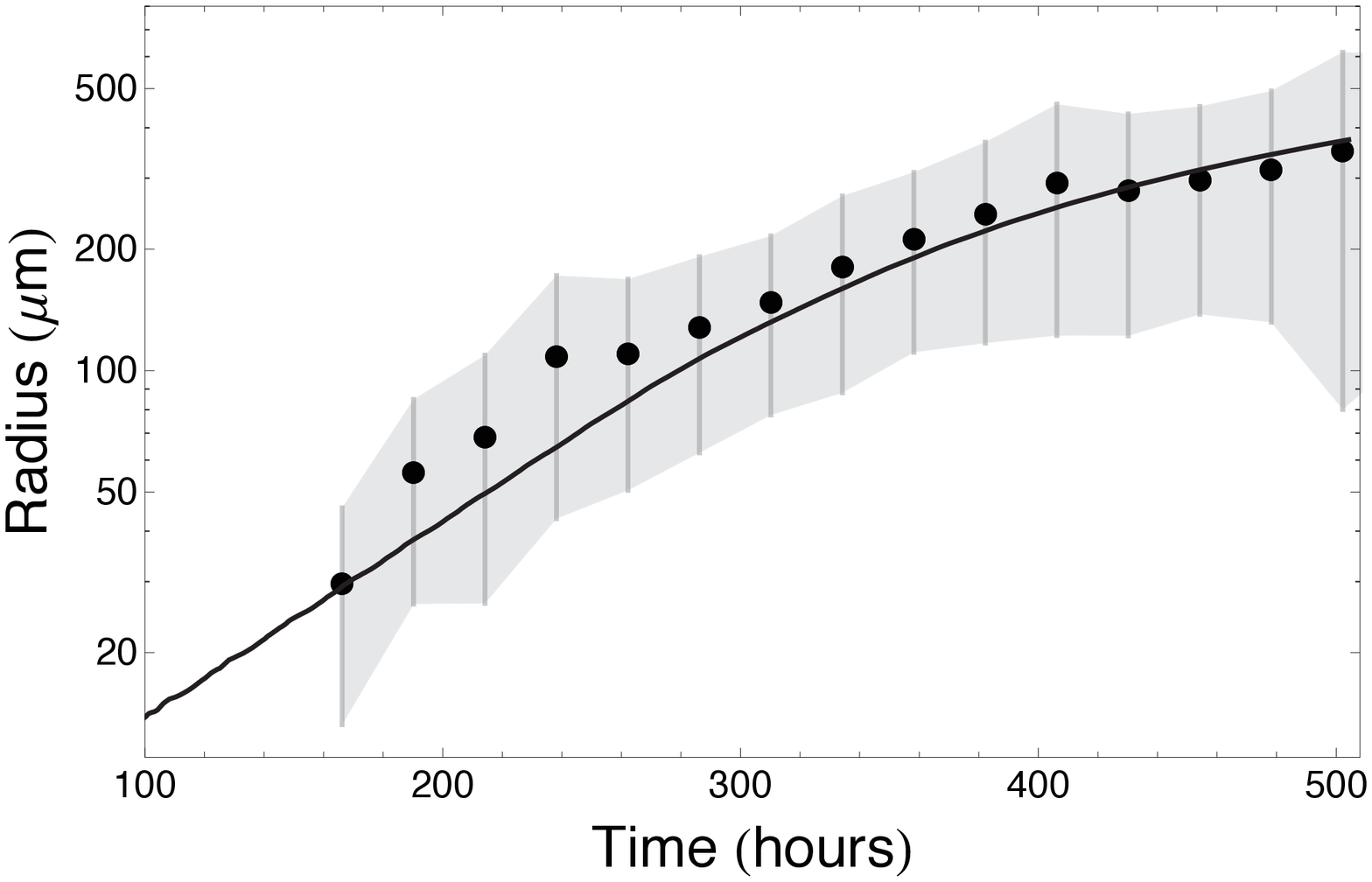,width=10cm}}
\vspace*{8pt}
\caption{Growth kinetics of a simulated muticell tumor spheroid and comparison with experimental data. The simulation (black line) was started by placing one virtual  cell in an environment of 1 ml filled with standard medium. The initial concentrations of glucose, glutamine, oxygen and the pH were set at the same values as for real experiments with tumor cells grown in a CO2 incubator and in RPMI medium supplemented with 10 $\%$ fetal bovine serum. The data in the figure (black symbols) are the mean $\pm$ standard deviation radii of nine spheroids obtained with 9L cells (rat glioblastoma cell line). Spheroids were grown isolated in individual culture wells under standard conditions, and their size was measured daily under the microscope (this experimental set was described in reference 42).}
\end{figure}

\section{Discussion and Conclusions}

\subsection{Parameter count}
\noindent
We model avascular tumor spheroids starting from individual molecular reactions in single cells and we climb the ladder of complexity up to large multicellular clusters. As shown above, this implies that we must deal with very different space and time scales, with the latter spanning 12 orders of magnitude. This effort unavoidably translates into a complex mathematical description that involves a vast number of equations and parameters, and with many free parameters one can in principle simulate any number of different patterns. Thus, a central question is whether our model can indeed be considered a realistic description of metabolism and mechanical evolution of cell clusters.
On the whole the present simulation is defined by 100 parameter values, and a realistic simulation requires a reliable set of parameters: to this end we extensively searched the scientific literature to find experimental measurements for as many parameters as possible. Whenever experimental measurements are missing, we assume values estimated from independent biophysical modeling of experimental data. Once fixed, the parameter values are not changed any more and simulation outputs are compared with new sets of experimental data to test the predictive power of the model itself (as we have shown here for cells growing in a bioreactor). All comparisons between simulation outputs and experimental data are finally carried out on a strict quantitative basis. Thus we feel confident that our numerical simulator is indeed a reliable model of the growth of tumor cells and tumor cell clusters.
Interestingly, we found that most parameters are correlated and can by no means assume arbitrary values. This is mainly due to the strong feedback between cells and the environment where simulated cells grow and that we have modeled. For example, one might tune metabolic parameters to allow cells to utilize nutrients more efficiently, produce energy under the form of ATP and grow faster. But this consequently results in a higher production and secretion of waste molecules that increase rapidly the acidity of the medium thus leading to cell death. Many parameters of the metabolic network of the cell are also strongly correlated, and this is a typical consequence of the interconnections between reactions that utilize different substrates for the same purpose, such as in the case of ATP and glutamine for both protein and DNA synthesis. As a consequence, the actual size of the parameter space is greatly reduced, and parameter tuning is far less complex than it appears to be at first sight. 

\subsection{A real-life simulation}
\noindent
Here we have reviewed the development of a numerical tool to simulate realistically the growth of avascular tumors and thus to explore the initial growth phase of solid tumors. This kind of numerical simulation has several important implications:
\begin{romanlist}[(ii)]
 \item it is possible to perform virtual experiments {\it in silico} that complement {\it in vitro} measurements, where many parameters are not directly accessible, and also {\it in vivo} observations, where accessibility problems are even greater also because of ethical issues. Our simulation program is indeed a virtual laboratory where one can make experiments at will in due time, and we hope that in the near future it will drive experimentalists towards the search of yet unexplored biological properties of tumors;
  \item the simulation focuses the modeling effort on the important details of cellular biophysics and spawns new ideas, both theoretical and experimental. For example, one important aspect that we have investigated to test the validity of our simulation program is whether, and eventually under which conditions, it could simulate a cell population with desynchronized cell cycles as it is observed for real cells.\cite{r10,r11} This prompted us to consider the relevant sources of internal randomness in cells, and in the attempt to investigate the biological causes of cell cycle desynchronization we developed theoretical tools  and carried out new experimental observations;\cite{r24,r25}
  \item the numerical model includes many complex non-linear interactions between different parts of the cell, and thus it has interesting predictive properties as unexpected biological behaviors can emerge;
  \item the model integrates several parts of our knowledge of cell biology, a knowledge that is fragmented in a huge number of small pieces throughout the scientific literature. An important aspect of our effort is that we are trying to connect together at least part of these pieces, and check their overall consistence. The model produces results that compare favorably with experimental data, and this indicates that it is possible to understand the cells' functions at the systemic level in quantitative terms;
  \item because of its incremental structure, our simulation program may serve as a platform to test the validity of other models of specific biochemical circuits. 
\end{romanlist}

For these reasons, we believe that our effort is by no means just a modeling exercise, but a serious and novel attempt to model cell biology, and a bridge between the molecular level -- albeit often phenomenological -- and the macroscopic observations, that demonstrates the possibility of carrying out realistic simulations and the evaluation of macroscopic parameters from microscopic descriptions.


\section*{Acknowledgments}
This work is supported by grants from the Istituto Nazionale di Fisica Nucleare, Group V, experiment VBL-RAD.


\end{document}